\documentclass[pra, aps, reprint, graphics, floatfix, tightenlines, twocolumn, longbibliography, superscriptaddress]{revtex4-1}

\pdfoutput=1

\usepackage{epsfig}
\usepackage{amsmath,amssymb,amsfonts}
\usepackage{mathrsfs}
\usepackage{bbm}
\usepackage{slashed}
\usepackage{graphicx}
\usepackage{verbatim}
\usepackage[usenames]{color}
\usepackage{mathtools}
\usepackage[ruled,linesnumbered,commentsnumbered]{algorithm2e}
\usepackage{float}
\usepackage{appendix}
\usepackage{booktabs}
\usepackage{relsize}
\usepackage{eurosym}
\usepackage{comment}
\usepackage{braket}
\usepackage{units}

\usepackage{enumerate}

\usepackage{tikz}

\usepackage{url}
\usepackage{hyperref}
\hypersetup{
    colorlinks=true,
    linkcolor=blue,
    citecolor=magenta,
    filecolor=magenta,
    urlcolor=cyan,
}

\newcommand{\sx}{\sigma^x}

\newcommand{\figref}[1]{\mbox{Fig.~\ref{#1}}}
\newcommand{\tabref}[1]{\mbox{Table~\ref{#1}}}
\newcommand{\secref}[1]{\mbox{Section~\ref{#1}}}

\renewcommand{\eqref}[1]{\mbox{Eq.~(\ref{#1})}}

\newcommand{\be}{\begin{equation}}
\newcommand{\ee}{\end{equation}}

\newcommand{\nn}{\nonumber}

\newcommand{\figpanel}[2]{Fig.~\hyperref[#1]{\ref*{#1}(#2)}}
\newcommand{\figpanels}[3]{Fig.~\hyperref[#1]{\ref*{#1}(#2)-(#3)}}
\newcommand{\figpanelNoPrefix}[2]{\hyperref[#1]{\ref*{#1}(#2)}}

\makeatletter
\def\@bibdataout@aps{%
\immediate\write\@bibdataout{%
@CONTROL{%
apsrev41Control%
\longbibliography@sw{%
    ,author="08",editor="1",pages="1",title="0",year="1"%
    }{%
    ,author="08",editor="1",pages="1",title="",year="1"%
    }%
  }%
}%
\if@filesw \immediate \write \@auxout {\string \citation {apsrev41Control}}\fi
}
\makeatother

\usepackage{mleftright} 

\usepackage[nolist]{acronym}

\begin{document}

\begin{acronym}
    \acro{ML}{machine learning}
\end{acronym}

\newacro{AQO}{adiabatic quantum optimization}
\newacro{HVRP}{heterogeneous vehicle routing problem}
\newacro{CVRP}{capacitated vehicle routing problem}
\newacro{VRP}{vehicle routing problem}
\newacro{TSP}{travelling salesperson problem}
\newacro{QAOA}{quantum approximate optimization algorithm}
\newacro{VQA}{variational quantum algorithm}
\newacro{VQAs}{variational quantum algorithms}
\newacro{NISQ}{noisy intermediate-scale quantum}
\newacro{BFGS}{Broyden–Fletcher–Goldfarb–Shanno}


\title{Applying quantum approximate optimization to \\ the heterogeneous vehicle routing problem}

\author{David Fitzek}
\email[]{davidfi@chalmers.se}
\affiliation{Department of Microtechnology and Nanoscience, Chalmers University of Technology, 412 96 Gothenburg, Sweden}
\affiliation{Volvo Group Trucks Technology, 405 08 Gothenburg, Sweden}

\author{Toheed Ghandriz}
\affiliation{Volvo Group Trucks Technology, 405 08 Gothenburg, Sweden}
\affiliation{Department of Mechanics and Maritime Sciences, Chalmers University of Technology, 412 96 Gothenburg, Sweden}

\author{Leo Laine}
\affiliation{Volvo Group Trucks Technology, 405 08 Gothenburg, Sweden}
\affiliation{Department of Mechanics and Maritime Sciences, Chalmers University of Technology, 412 96 Gothenburg, Sweden}

\author{Mats Granath}
\email[]{mats.granath@physics.gu.se}
\affiliation{Department of Physics, University of Gothenburg, 412 96 Gothenburg, Sweden}

\author{Anton Frisk Kockum}
\email[]{anton.frisk.kockum@chalmers.se}
\affiliation{Department of Microtechnology and Nanoscience, Chalmers University of Technology, 412 96 Gothenburg, Sweden}


\begin{abstract}
Quantum computing offers new heuristics for combinatorial problems. With small- and intermediate-scale quantum devices becoming available, it is possible to implement and test these heuristics on small-size problems.
A candidate for such combinatorial problems is the \ac{HVRP}: the problem of finding the optimal set of routes, given a heterogeneous fleet of vehicles with varying loading capacities,  to deliver goods to a given set of customers.
In this work, we investigate the potential use of a quantum computer to find approximate solutions to the \ac{HVRP} using the quantum approximate optimization algorithm (QAOA).
For this purpose we formulate a mapping of the \ac{HVRP} to an Ising Hamiltonian and simulate the algorithm on problem instances of up to 21 qubits. We find that the number of qubits needed for this mapping scales quadratically with the number of customers. We compare the performance of different classical optimizers in the QAOA for varying problem size of the \ac{HVRP}, finding a trade-off between optimizer performance and runtime.

\end{abstract}

\maketitle


\section{Introduction}

Devices utilizing quantum-mechanical effects provide a new computational paradigm that enables novel algorithms and heuristics~\cite{Shor2002, Harrow2009, Montanaro2016, Wendin2017, Bharti2021}. The ongoing development of such devices~\cite{Arute2019, Bruzewicz2019, Kjaergaard2020, Zhong2020} provides an opportunity to test these algorithms on small problem instances, which could lead to new solutions to hard optimization problems.
In this work, we show how a \ac{QAOA}~\cite{Farhi2014} can be employed to find approximate solutions for the heterogeneous vehicle routing problem (HVRP)~\cite{Koc2016}. Our approach can be utilized on both \ac{NISQ}~\cite{Preskill2018} computers and quantum annealers~\cite{Hauke2020}. It also paves the way for implementing challenging instances of the \ac{HVRP} on larger quantum computers in the future. 

The \ac{HVRP} belongs to the well known and extensively studied class of optimization problems known as the \ac{VRP}~\cite{Golden2008} in the field of logistics. The \ac{VRP} captures the problem of finding the optimal set of routes for a fleet of vehicles to travel in order to deliver goods to a given set of customers. This problem is also found in supply-chain management and scheduling~\cite{Caunhye2012}. Variants of the \ac{VRP} include the \ac{CVRP}, in which the vehicles have a limited carrying capacity~\cite{Golden2008}, and the \ac{HVRP} studied here, in which the fleet composition is unknown and capacity constraints are given~\cite{Koc2016, Ghandriz2021}. All these VRPs are very challenging since they belong to the complexity class NP-hard~\cite{Lenstra1981}. 

Due to its industrial relevance, there has been tremendous effort devoted to finding good approximate solutions to the \ac{VRP} and its variants through various heuristics~\cite{Laporte2014, Tavares2009}, e.g., construction heuristics~\cite{Hwang1999}, improvement heuristics~\cite{VanBreedam1995}, and metaheuristic top-level strategies~\cite{Brandstatter2018}. In construction heuristics, e.g., the Clarke and Wright saving algorithm~\cite{Clarke1964}, one starts from an empty solution and iteratively extends it until a complete solution is obtained. In improvement heuristics, one instead starts from a complete solution (often generated by a construction heuristic) and then try to improve further through local moves. There are several software libraries and tools that implement ready-to-use solvers for all these methods~\cite{Loungee2003, Groer2010, ortools}. Moreover, exact methods for solving the \ac{VRP} and its variants have also been investigated~\cite{Baldacci2009}.

In this article, we instead investigate a heuristic method for solving the \ac{HVRP} on a quantum computer. Such devices, including both programmable quantum processors~\cite{Wendin2017, Jazaeri2019, Kjaergaard2020} and quantum annealers~\cite{Hauke2020}, are gradually becoming available due to the recent advances in controlling quantum systems. The current quantum computers are known as \ac{NISQ} devices~\cite{Preskill2018}, since they are largely limited by their intermediate number (several tens~\cite{Arute2019, Pino2021, Mooney2021, Blinov2021, Pogorelov2021a, Wu2021}) of controllable qubits, limited connectivity, imperfect qubit control, short coherence times, and minimal error correction. Only a subset of known quantum algorithms can run on these near-term devices~\cite{Kandala2017, Bharti2021}; other algorithms require more advanced hardware. 

The heuristic method we apply to the \ac{HVRP} here is an example of a \ac{VQA}~\cite{Cerezo}, which is a promising class of quantum algorithms that are compatible with \ac{NISQ} devices. 
These algorithms generally need access to a description of the problem, and also possibly to a set of training data. With this in hand, the first step is to define a cost (or loss) function $C$, which encodes the solution to the problem. Next, one proposes an ansatz, i.e., a quantum operation depending on a set of continuous or discrete parameters $\boldsymbol{\theta}$ that can be optimized. This ansatz is then optimized in a hybrid quantum-classical loop to solve the optimization task
\be
\boldsymbol{\theta^*} =\underset{\boldsymbol{\theta}}{\text{argmin }} C(\boldsymbol{\theta}).
\ee
Such algorithms have emerged as a leading contender for obtaining quantum advantage~\cite{Cerezo} within the constraints of \ac{NISQ} devices. By now, \ac{VQAs} have been proposed for numerous applications that researchers have envisioned for quantum computers, e.g., in chemistry, logistics, and finance~\cite{Stamatopoulos2019, Vikstal2019, Braine2019, Choi2019, Egger2020}.

The type of \ac{VQA} we employ here is the \ac{QAOA}~\cite{Farhi2014}, which is a heuristic that can approximate the solution to many combinatorial problems, including VRPs. Current research in this area ranges from applications on large-scale \ac{VRP} instances with a quantum annealer~\cite{Syrichas2017} to more specific variants of the \ac{VRP}, such as the \ac{CVRP}~\cite{Feld2019} or the multi-depot capacitated \ac{VRP}~\cite{Harikrishnafkumar2020}. These approximation algorithms have been tested on quantum annealers~\cite{Feld2019} and \ac{NISQ} devices~\cite{Utkarsh2020}. There have also been several experimental realizations of the \ac{QAOA} applied to other optimization problems~\cite{Harrigan,Bengtsson2020, Abrams2020, Willsch2020, Otterbach2017}.

However, a problem description suited for the \ac{QAOA}, an Ising Hamiltonian~\cite{Ising1925, Brush1967, Lucas14} (describing the energy of interacting two-level systems), seems to be lacking for the case of the \ac{HVRP}. In this work, we provide such a mapping for the \ac{HVRP}, which can be utilized on both \ac{NISQ} computers and quantum annealers. We show that, in this formulation, the number of qubits scales quadratically with the number of customers. To explore the performance of the QAOA applied to the HVRP, we simulate problem instances with up to 21 qubits. We check how this performance depends both on the choice of classical optimizer and on the depth of the quantum circuit. This work lays the foundation for finding approximate solutions to large problem instances of the \ac{HVRP} when sufficiently advanced quantum-computing hardware becomes available.

The paper is organized as follows. In Sec.~\ref{sec:heterogeneous_vehicle_routing_problem}, we introduce the \ac{HVRP} and its mathematical formulation. Then, in Sec.~\ref{sec:ising_formulation_for_the_hvrp}, we develop the Ising formulation of the \ac{HVRP}. In Sec.~\ref{sec:qaoa}, we review the \ac{QAOA} and describe how it can be used to find approximate solutions to the \ac{HVRP}. In Sec.~\ref{sec:approximate_solution_for_the_hvrp}, we present numerical results from applying the \ac{QAOA} to a few HVRPs of different sizes. Finally, we conclude the paper and give an outlook for future work in Sec.~\ref{sec:conclusion}.


\section{The heterogeneous vehicle routing problem}
\label{sec:heterogeneous_vehicle_routing_problem}

The \ac{HVRP} can be formulated as follows~\cite{Koc2016}. A fleet of vehicles is available at a depot, which becomes node 0 of a complete graph $\mathcal{G} = (\mathcal{N}, \mathcal{E})$ (we do not consider multiple depots). Here, $\mathcal{N} = \{ 0,  ... , n \} $ is the set of nodes or vertices, such that the $n$ customers that the fleet of vehicles should deliver goods to constitute the customer set $\mathcal{N}_0 = \mathcal{N} \setminus \{0\}$, and $\mathcal{E} = \{(i, j) :0 \le i, j \le n, i \ne j\}$ denotes the set of edges or arcs. Each customer $i$ has a positive demand $q_i$.

The set of available vehicle types is $\mathcal{V} = \{1,...,k \}$, with $m_v$ vehicles of type $v \in \mathcal{V}$. When using these vehicles to deliver goods to meet the customer demand, there are several costs and constraints that need to be taken into account. First, there is the fixed vehicle cost $t^v$, i.e., the cost that is independent of the distance travelled by the vehicle of type $v$. Then, there is the vehicle capacity $Q^v$. Note that different vehicle types can have the same capacities, but differ in, e.g., the type of powertrain used~\cite{Ghandriz2021}. Finally, there is the cost $c_{ij}^v$ of travelling on edge $(i,j)$ with the vehicle of type $v$. To describe all constraints, it is also useful to introduce the binary variables $x^v_{ij}$, which are equal to 1 if and only if a vehicle of type $v$ travels on edge $(i, j)$. Furthermore, we denote by $f_{ij}^v$ the amount of goods that are leaving node $i$ to go to node $j$ using truck $v$, while the amount of goods entering the node is denoted $f_{ji}^v$.

Using this notation, the HVRP is to minimize the cost
\be
C_{\rm tot} = \sum_{v \in \mathcal{V}} \sum_{j \in \mathcal{N}_0} t^v x^v_{0j} + \sum_{v \in \mathcal{V}} \sum_{(i,j) \in \mathcal{E}} c^v_{ij} x^v_{ij} \, ,
\label{equ:min}
\ee
subject to the constraints
\begin{align}
&\sum_{j \in \mathcal{N}_0} x^v_{0j} \le m_v \quad v\in \mathcal{V} \, , \label{equ:sub} \\
&\sum_{v \in \mathcal{V}} \sum_{j \in \mathcal{N}} x^v_{ij} = 1 \quad i \in \mathcal{N}_0 \, , \label{equ:3} \\
&\sum_{v \in \mathcal{V}} \sum_{i \in \mathcal{N}} x^v_{ij} = 1 \quad j \in \mathcal{N}_0 \, , \label{equ:4} \\
&\sum_{j \in \mathcal{N}_0} x^v_{j0} = \sum_{j \in \mathcal{N}_0} x^v_{0j} \quad v\in \mathcal{V} \, , \label{equ:5} \\
&\sum_{v \in \mathcal{V}} \sum_{j \in \mathcal{N}} f^v_{ji} - \sum_{v \in \mathcal{V}} \sum_{j \in \mathcal{N}} f^v_{ij} = q_i \quad i \in \mathcal{N}_0 \, , \label{equ:6} \\
&q_j x_{ij}^v \le f^v_{ij} \le (Q_v - q_i)x_{ij}^v \quad (i,j) \in \mathcal{E}, v \in \mathcal{V} \, , \label{equ:7} \\
&x_{ij}^v \in \{0,1\} \quad (i,j) \in \mathcal{E}, v \in \mathcal{V} \, , \label{equ:8} \\
&f^v_{ij} \ge 0 \quad (i,j) \in \mathcal{E}, v \in \mathcal{V} \, . \label{equ:9}
\end{align}
The objective function in \eqref{equ:min} is the sum of the fixed vehicle cost for the vehicles used to deliver goods and the total (variable) travel cost for those vehicles. The constraint in \eqref{equ:sub} ensures that the maximum number of available vehicles for a specific vehicle type is not exceeded. The constraints in Eqs.~(\ref{equ:3}) and (\ref{equ:4}) make sure that each customer is visited exactly once, and the constraint in \eqref{equ:5} sees to that all vehicles leaving the depot return to it after delivering their goods. The constraints in Eqs.~(\ref{equ:6}) and (\ref{equ:7}) ensure a correct commodity flow that meets all customer demands. Finally, the constraints in Eqs.~(\ref{equ:8}) and (\ref{equ:9}) enforce the binary form and non-negativity restrictions on the variables.


\section{Ising formulation for the \ac{HVRP}}
\label{sec:ising_formulation_for_the_hvrp}

All optimization problems in the complexity class NP can be reformulated as the problem of finding the ground state (lowest-energy configuration) of a quantum Hamiltonian~\cite{Lucas14}. This is also the method we use for the \ac{HVRP} in this work. Since the \ac{HVRP} combines two distinct problems, a routing problem and a capacity problem, we have to derive an Ising Hamiltonian that captures both these problems simultaneously.


\subsection{Routing problem}

For the routing problem, we start from the \ac{TSP} formulation given in Ref.~\cite{Lucas14} with the Hamiltonian $H$ encoding the total cost:
\begin{align}
    H &= H_A + H_B \label{equ:tsp} \, , \\
    H_A &= A \sum_{i=1}^N \mleft( 1 - \sum_{\alpha=1}^N y_{i \alpha} \mright)^2 + A \sum_{\alpha=1}^N \mleft(1 - \sum_{i=1}^N y_{i \alpha} \mright)^2 \nn \\
        &+ A \sum_{(i, j) \notin \mathcal{E}} \sum_{\alpha=1}^N y_{i \alpha} y_{j \alpha+1} \, , \label{equ:tsp_ha} \\
    H_B &= B \sum_{(i, j) \in \mathcal{E}} W_{i j} \sum_{\alpha=1}^N y_{i \alpha} y_{j \alpha+1} \, , \label{equ:tsp_hb}
\end{align}
where $N = |\mathcal{N}|$ is the number of nodes including the depot, $A$ and $B$ are positive constants, and $W$ encodes the distances between the nodes. The index $i$ represents the nodes and $\alpha$ the order in a prospective cycle. The binary variables $y_{i \alpha}$ can be referred to as 'routing variables' indicating in which order of the cycle node $i$ is visited. There are $N^2$ variables, with $y_{i,N+1}\equiv y_{i,1}$ for all $i$, such that the route ends where it starts. The last term in \eqref{equ:tsp_ha}, which ensures that non-existent edges are not used, can be neglected for the problem we investigate because we assume a fully connected graph.

To combine this formulation with the mathematical formulation of the \ac{HVRP} given in Eqs.~(\ref{equ:min})--(\ref{equ:9}), we need a map from the decision variable $y$ to $x$. The map we use is
\begin{align}
    x_{ij}^v &= \sum_{\alpha=1}^{N_0-1} y_{i\alpha}^v y_{j\alpha+1}^v \, , \label{equ:x_ij} \\ 
    x_{0i}^v &= y_{i1}^v + \sum_{\alpha=2}^{N_0} \mleft( 1 - \sum_{\substack{j=1 \\ j \ne i}}^{N_0} y_{j \alpha-1}^v \mright) y_{i \alpha}^v \, , \label{equ:x_0i} \\
    x_{i0}^v &= y_{i N_0}^v + \sum_{\alpha=1}^{N_0 - 1} y_{i \alpha}^v \mleft( 1 - \sum_{\substack{j=1 \\ j \ne i}}^{N_0} y_{j \alpha+1}^v \mright) \, . \label{equ:x_i0}
\end{align}
The summation in \eqref{equ:x_ij} is not equal 0 if and only if $i$ and $j$ are subsequent stops on the same route. Equations~(\ref{equ:x_0i}) and (\ref{equ:x_i0}) ensure that the first and last stops are automatically connected to the depot (assuming a single depot). Remember that the index 0 denotes the depot and the index 1 the first city (node) in the list of cities (nodes).

We can now write the Ising formulation for the routing problem. Here we extend the formulation compared to previous works to capture different types of trucks, not just multiple trucks of the same type (having the same capacity)~\cite{Lucas14, Roch2020}. Let $V = |\mathcal{V}|$ be the number of trucks, where $\mathcal{V}$ now is the set of vehicles chosen for the optimization (instead of the set of vehicle \textit{types}, as in \secref{sec:heterogeneous_vehicle_routing_problem}), and denote by $N_0 = |\mathcal{N}_0|$ the number of customers to visit. The indices $v$ now represent a specific truck of a specific type (instead of just a specific type, as in \secref{sec:heterogeneous_vehicle_routing_problem}). The Ising Hamiltonian we derive is then
\begin{align}
    H &= H_A + H_B + H_C + H_D \, , \label{equ:hamiltonian} \\
    H_{A} &= A \sum_{v=1}^V \sum_{i=1}^{N_0} \sum_{j=1}^{N_0} c^v_{ij} \sum_{\alpha=1}^{N_0 - 1} y_{i \alpha}^v y_{j \alpha + 1}^v \nn \\
        &+ A \sum_{v=1}^V \sum_{i=1}^{N_0} c_{0i}^v \mleft[ y_{i1}^v + \sum_{\alpha=2}^{N_0} \mleft( 1 - \sum_{\substack{j=1 \\ j \ne i}}^{N_0} y_{j \alpha-1}^v \mright) y_{i \alpha}^v \mright] \nn \\
        &+ A \sum_{v=1}^V \sum_{i=1}^{N_0} c_{i0}^v \mleft[ y_{i N_0}^v + \sum_{\alpha=1}^{N_0 - 1} y_{i \alpha}^v \mleft(1 - \sum_{\substack{j=1 \\ j \ne i}}^{N_0} y_{j \alpha+1}^v \mright) \mright] \, , \label{equ:h_a_routing} \\
    H_B &= B \sum_{j=1}^{N_0} \sum_{v=1}^V t^v \sum_{\alpha=2}^{N_0} \mleft(1 - \sum_{i=1}^{N_0} y_{i \alpha-1}^v \mright) y_{j \alpha}^v \, , \label{equ:h_b_routing} \\
    H_C &= C \sum_{i=1}^{N_0} \mleft(1 - \sum_{\alpha=1}^{N_0} \sum_{v=1}^V y_{i \alpha}^v \mright)^2 \, , \label{equ:h_c_routing} \\
    H_D &= D \sum_{\alpha=1}^{N_0} \mleft(1- \sum_{i=1}^{N_0} \sum_{v=1}^V y_{i\alpha}^v \mright)^2 \, . \label{equ:h_d_routing}
\end{align}

The Hamiltonian $H$ in \eqref{equ:hamiltonian} is composed of different parts. Here, $H_A$ in \eqref{equ:h_a_routing} captures the first part of the original mathematical formulation, i.e., the minimization of the variable cost. The first term estimates the variable cost for traveling between the different customers/cities, while the second and third terms measure the cost of leaving and arriving at the depot.
For this particular mapping it is necessary to define the set of vehicles that are used for the optimization beforehand. Therefore, we can neglect the inequality constraint defined in \eqref{equ:sub} from the original formulation, which ensures that the number of vehicles of a specific type does not exceed the number of available vehicles.
Similarly, $H_B$ in \eqref{equ:h_b_routing} estimates the fixed costs of each vehicle leaving the depot [see \eqref{equ:min}]. Note that the prefactors $A$ and $B$ must be equal, in order not to rescale the relative fixed versus variable costs. The constraint given by $H_C$ in \eqref{equ:h_c_routing} ensures that each city is visited exactly once. Furthermore,
$H_D$ in \eqref{equ:h_d_routing} ensures that each city has a unique position in the cycle and that not more than one city can be travelled to at the same time. To make sure that the constraints are not violated, we require $0 < \text{max}(H_A + H_B) < C, D$.

\begin{figure}
    \centering
    \includegraphics[width=0.95\linewidth]{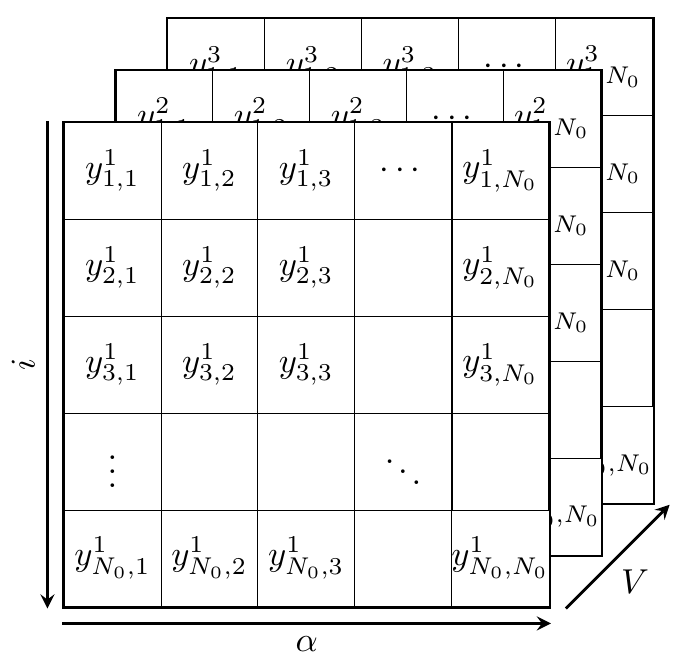}
    \caption{Visualization of the decision variables $y_{ij}^v$ in the Ising formulation of the routing problem.
    \label{fig:decision_vars}}
\end{figure}

The decision variables $y_{ij}^v$ are positioned as shown in \figref{fig:decision_vars}. This picture allows us to see the operations that are taking place when summing over specific indices. As an example, consider \eqref{equ:h_c_routing}. First summing over the indices $v$ and $\alpha$ corresponds to a summation over these two axes. After the summation it is easy to see that if the goal is to visit each customer/city once, then each element of the length $N_0$ array must be one.

\begin{figure}
    \centering
    \includegraphics[width=\linewidth]{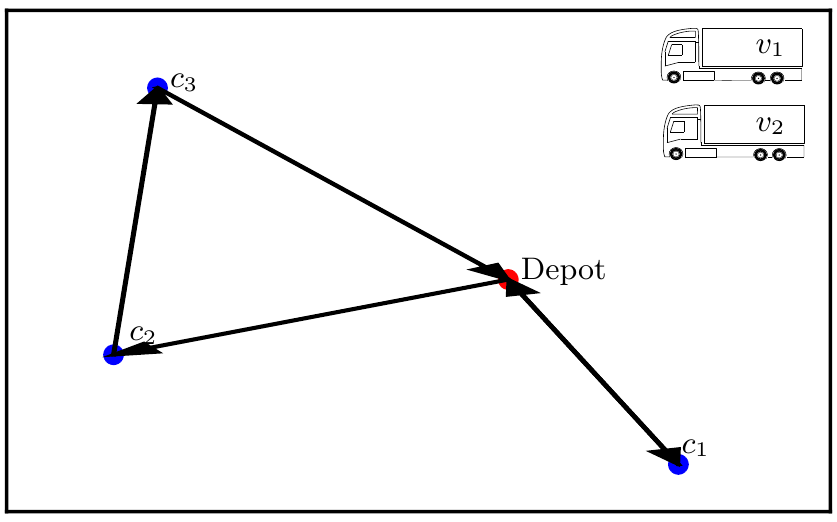}
    \caption{Visualization of a problem instance with a suggested solution. The first truck $v_1$ visits city $c_2$ and the city $c_3$ before returning to the depot. The second truck $v_2$ only visits city $c_1$.}
    \label{fig:instance}
\end{figure}
One notable technicality about the formulation is that certain solutions that may be considered valid are excluded by the constraint in \eqref{equ:h_d_routing}. However, the excluded solutions are physically equivalent to some allowed solution, as illustrated by the following simple example (see \figref{fig:instance}) with two trucks over three cities
\be
    y =
    \mleft[ \begin{pmatrix}
        0 & 0 & 0 \\
        1 & 0 & 0 \\
        0 & 1 & 0
    \end{pmatrix},
    \begin{pmatrix}
        0 & 0 & 1 \\
        0 & 0 & 0 \\
        0 & 0 & 0
    \end{pmatrix} \mright]\,,\label{equ:sol_2}
\ee
and
\be
    y =
    \mleft[ \begin{pmatrix}
        0 & 0 & 0 \\
        1 & 0 & 0 \\
        0 & 1 & 0
    \end{pmatrix},
    \begin{pmatrix}
        0 & 1 & 0 \\
        0 & 0 & 0 \\
        0 & 0 & 0
    \end{pmatrix} \mright]\,. \label{equ:sol_1}
\ee
In both cases, the two different matrices describe the routes of the two different vehicles and the order in which they visit the different customers $c_i$. They both represent a physically valid solution where the first truck visits first the second and then the third customer ($v_1: c_2 \rightarrow c_3$), while the second truck goes from the depot to the first customer and then back to the depot ($v_2: c_1 $). The constraint in \eqref{equ:h_d_routing}, however, rules out the latter representation as it has two non-zero entries for $\alpha=2$. If we want to allow this larger set of viable representations, physically equivalent to solutions already allowed by \eqref{equ:h_d_routing}, we can replace that constraint by a reformulated one,
\be
    H'_D = D \sum_{v=1}^V \sum_{\alpha=1}^{N_0} \mleft(u_\alpha^v - \sum_{i=1}^{N_0} y_{i \alpha}^v  \mright)^2\,,
    \label{equ:h_d_alternative}
\ee
where we have introduced $N_0^2$ additional auxiliary qubits $u_{\alpha}^v$. Especially in the \ac{NISQ} era, where quantum resources are scarce, it is important to encode the problem with as few qubits as possible. Thus we do not consider \eqref{equ:h_d_alternative} a viable route for implementations, but use \eqref{equ:h_d_routing} for the simulations in \secref{sec:approximate_solution_for_the_hvrp}.


\subsection{Capacity problem}

The capacity constraint is of a similar nature as the constraints for the knapsack problem --- both are described by an inequality constraint, which for the knapsack problem is to not add too many items to the knapsack and for the trucks to not overload the vehicles. Therefore, we can use the formulation given in Ref.~\cite{Lucas14} to model the inequality constraint introduced by the capacities.

The knapsack problem with integer weights is the following. We have a list of $N$ objects, labeled by $i$, with the weight of each object given by $w_i$ and its value by $c_i$. The knapsack has a limited capacity of $W$. The binary decision variable $x_i$ denotes whether an item is contained (1) in the knapsack or not (0). The total weight of the knapsack is
\be
    \mathcal{W} = \sum_{i=1}^N w_i x_i
\ee
with a total value of
\be
    \mathcal{C} = \sum_{i=1}^N c_i x_i \, .
\ee
The NP-hard knapsack problem is to maximize $\mathcal{C}$ while satisfying the inequality constraint $\mathcal{W} \le W$.

We can write an Ising formulation of the knapsack problem as follows. Let $z_n$ for $1 \le n \le W$ be a binary variable which is 1 if the final weight of the knapsack is $n$ and 0 otherwise~\cite{Lucas14}. The Hamiltonian whose energy we seek to minimize is then
\begin{align}
    H &= H_A + H_B \, , \\
    H_A &= A \mleft( 1 - \sum_{n=1}^W z_n \mright)^2 + A \mleft( \sum_{n=1}^W n z_n - \sum_{i=1}^N w_i x_i \mright)^2 \, , \label{equ:knapsack_constraint} \\
    H_B &= -B \sum_{i=1}^N c_i x_i \label{equ:knapsack_optimization} \, .
\end{align}
To make sure that the hard constraint is not violated, we require $0 < \text{max}(|H_B|) < A$. 


\subsubsection{Reducing the number of auxiliary qubits}
\label{reduce_number_of_auxiliary_qubits}

It is possible to reduce the number of variables required for the auxiliary variable $z_n$. We want to encode a variable which can take the values from 0 to $W$. Let $M \equiv \lfloor \text{log}_2(N) \rfloor$. We then require $M+1$ binary variables instead of $N$ binary variables:
\be
    \sum_{n=1}^N n z_n \rightarrow \sum_{n=0}^{M-1} 2^n z_n + \mleft( N + 1 - 2^M \mright) z_M \, .
\ee
Note that if $ N \ne 2^{M+1} -1 $, degeneracies are possible~\cite{Lucas14}. Within this log formulation, several of the auxiliary variables can be 1, so the first part of \eqref{equ:knapsack_constraint} should not be included as this constraint enforces a one-hot encoding (exactly one element of the bitstring is one and the rest are zero) of the bitstrings.

We can make use of the inequality constraint given in the knapsack formulation [see \eqref{equ:knapsack_constraint}] to encode the capacity constraints for the \ac{HVRP}. Therefore, we can neglect $H_B$ [see \eqref{equ:knapsack_optimization}] and only consider $H_A$ [see \eqref{equ:knapsack_constraint}]. Let $Q^v$ be the maximum capacity of vehicle $v$. The Hamiltonian then becomes
\begin{align}
    H_A &= A \sum_v \mleft(1 - \sum_{k=0}^{Q^v} z_k^v \mright)^2 + A \sum_v \mleft( \sum_{k=0}^{Q^v} k \cdot z_k^v - \sum_{\alpha,i} q_i y_{i \alpha}^v \mright)^2,
\end{align}
or equivalently using the log formulation,
\begin{align}
    H_A &= A \sum_v \mleft( \sum_{k=0}^{M^v-1} 2^k z_k^v + (Q^v + 1 - 2^{M^v}) z_{M^v}^v - \sum_{\alpha,i} q_i y_{i \alpha}^v \mright)^2.
\end{align}
Note that by using the log trick, the decision variable $z_k^v$ switches from a one-hot encoding to a binary representation. 


\subsection{The full Ising Hamiltonian for the \ac{HVRP}}

We are now ready to write down the full Hamiltonian for the \ac{HVRP}. The full Ising Hamiltonian $H_{\mathcal{C}}$ contains five terms, where the first term $H_A$ captures the actual optimization problem and the other terms are penalty terms to ensure that invalid configurations are penalized with a high energy:
\begin{align}
    H_{\mathcal{C}} &= H_A + H_B + H_C + H_D + H_E \, , \label{equ:cost_hamiltonian} \\
    H_E &= E \sum_{v=1}^V \mleft( \sum_{k=0}^{M^v-1} 2^k z_k^v + (Q^v + 1 - 2^{M^v}) z_{M^v}^v \mright. \nn \\
        & \mleft. - \sum_{\alpha=1}^{N_0} \sum_{i=1}^{N_0} q_i y_{i \alpha}^v \mright)^2 \label{equ:h_e_knapsack}
\end{align}
For the terms $H_A$ to $H_D$, see Eqs.~(\ref{equ:h_a_routing})--(\ref{equ:h_d_routing}).

The formulation presented in this paper combines the capacity problem and the routing problem in one Ising Hamiltonian. Similarly, a unified approach is also attempted in Ref.~\cite{Feld2019} with the difference that the authors add a constraint for clustering the customers as well. Here, by using the decision variables $y_{i \alpha}^v$ that indicate the position in a prospective cycle instead of $x^v$ that is 1 if and only if a vehicle traverses from customer $i$ to $j$, we circumvent the subtour-elimination constraint.
This constraint needs to loop through all possible subtours as it is presented in Ref.~\cite{Harikrishnafkumar2020}.
Moreover, a solution obtained with our mapping does not necessarily use all the vehicles that are available. It can find the most cost efficient subset of vehicles needed to solve the task.


\subsection{Resources}

The required resources (qubits) for solving the \ac{HVRP} with our approach can be separated into three parts. The first part comes from encoding the connections between the customers and scales with $N_0^2 \cdot V$. Additionally, auxiliary qubits are required for the constraining term $H_E$. The overall number of qubits, $\#q$, required are
\be
    \#q = N_0^2 \cdot V + \underbrace{\sum_{v=1}^{V} \lfloor \log_2 Q^v \rfloor + 1}_{H_E} \, .
\ee
Using the alternative formulation for $H_D$ [see \eqref{equ:h_d_alternative}] adds more auxiliary qubits ($N_0 \cdot V$), yielding
\be
    \#q =(N_0 + 1) N_0 V + \sum_{v=1}^{V} \lfloor \log_2 Q^v \rfloor + 1 \, .
\ee

As a comparison, we note that modern high-performance optimizers (classical computers) for the HVRP can solve problem instances with more than 1,000 customers~\cite{Uchoa2017a, Barman}. For a quantum computer to solve problem instances of this size, it would need at least millions of controllable qubits. Systems of this size are likely still several years away from being realized. 


\section{The quantum approximate optimization algorithm}
\label{sec:qaoa}

\begin{figure*}
    \centering
    \includegraphics[width=\linewidth]{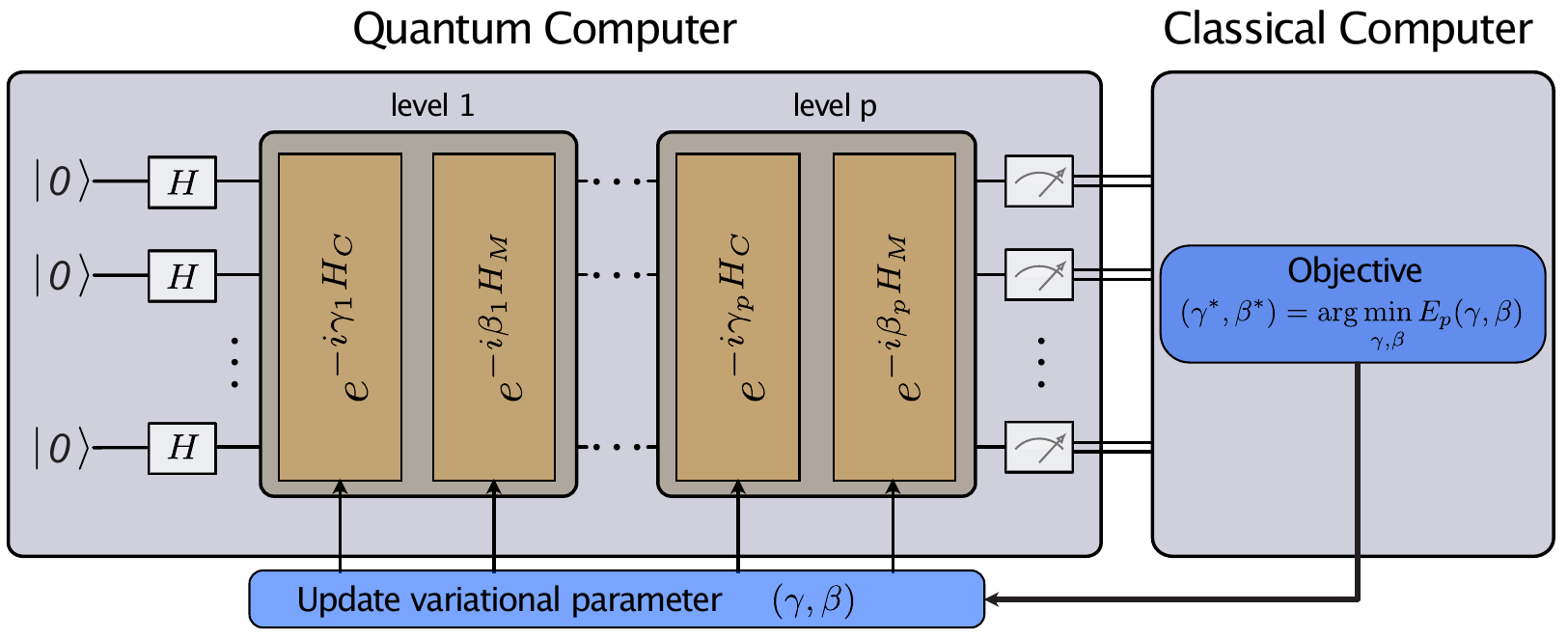}
    \caption{A schematic description of the \ac{QAOA}, visualizing the interplay between the quantum device and the classical computer. The quantum computer implements a variational state formed by applying $p$ parameterized layers of operations. Each layer has operations involving the cost Hamiltonian $H_C$ and a mixing Hamiltonian $H_M$, weighted by the angles $\gamma$ and $\beta$, respectively. Measurements of the variational state and calculations of its resulting energy are used to guide the classical optimizer, which minimizes the energy in a closed-loop optimization.}
    \label{fig:qaoa}
\end{figure*}

The \ac{QAOA} belongs to the class of hybrid quantum-classical algorithms, which combine quantum and classical processing. The closed-loop optimization of the classical and quantum devices is visualized in \figref{fig:qaoa}. The quantum subroutine, operating on $n$ qubits, consists of a consecutive application of two non-commuting operators defined as~\cite{Farhi2014}
\begin{align}
    U(\gamma) &\equiv e ^{-i\gamma H_\mathcal{C}} \quad \gamma \in [0, 2\pi] \, , \label{equ:h_cost} \\
    U(\beta) &\equiv e^{-i\beta H_M} = \prod_{j=1}^n e^{-i\beta \sx_j} \quad  \beta \in [0, \pi] \, , \label{equ:h_mixing} 
\end{align}
where $\sigma^x$ denotes the Pauli $X$ matrix~\cite{Nielsen2000}. This operation is analogous to the classical NOT gate. It changes the $\ket{0}$ state to the $\ket{1}$ state, and vice versa. The operator $U(\gamma)$ gives a phase rotation to each bit string depending on the cost of the string, while the mixing term $U(\beta)$ scrambles the bit strings. We call $H_\mathcal{C}$ the cost Hamiltonian and $H_M$ the mixing Hamiltonian. The bounds for $\gamma$ and $\beta$ are valid if $H_\mathcal{C}$ has integer eigenvalues~\cite{Farhi2014}. The formulation of $H_\mathcal{C}$ for the \ac{HVRP} is given by \eqref{equ:cost_hamiltonian}.

The initial state for the algorithm is a superposition of all possible computational basis states. This superposition can be obtained by first preparing the system in the initial state $\ket{0}^{\otimes n}=\ket{00 \ldots 0}$ for all qubits and then applying the Hadamard gate on each qubit:  
\begin{align}
    \mleft(\tilde{H}\ket{0}\mright)^{\otimes n} &= \mleft(\frac{\ket{0} + \ket{1}}{\sqrt{2}}\mright)^{\otimes n} \equiv \ket{+}^{\otimes n} \, ,
\end{align}
where $\otimes$ denotes the tensor product and $\tilde{H}$ the Hadamard gate.

For any integer $p \ge 1$ and $2p$ angles $\gamma_1 \dots \gamma_p \equiv \boldsymbol{\gamma} $ and $\beta_1 \dots \beta_p \equiv \boldsymbol{\beta}$, we define the angle-dependent quantum state
\be
    \ket{\boldsymbol{\gamma, \beta}} = U( \beta_p) U(\gamma_p) \dots U(\beta_1) U( \gamma_1) \ket{+}^{\otimes n}.
\ee
The quantum circuit parametrized by $\boldsymbol{\gamma}$ and $\boldsymbol{\beta}$ is then optimized in a closed loop using a classical optimizer. The objective is to minimize the expectation value of the cost Hamiltonian $H_{\mathcal{C}}$~\cite{Farhi2014}, i.e.,
\begin{align}
    (\boldsymbol{\gamma^*, \beta^*}) &=\underset{\boldsymbol{\gamma, \beta}}{\text{argmin }}  E(\boldsymbol{\gamma, \beta}) \, , \\
    E(\boldsymbol{\gamma, \beta}) &= \bra{\boldsymbol{\gamma, \beta}} H_{\mathcal{C}} \ket{\boldsymbol{\gamma, \beta}} \, .
\end{align}
The problem of calculating the energy of $2^{\#q}$ possible bit strings (solutions) is thus reduced to a variational optimization over $2p$ parameters.


\section{Benchmarking quantum approximate optimization for the heterogeneous vehicle routing problem}
\label{sec:approximate_solution_for_the_hvrp}

\begin{table}
    \caption{Information about the three different problem instances used in simulations.}
    \label{tab:problem_instance}
    \begin{tabular}{lcccl}
        \hline
        \hline
        Problem instance & I & II & III &  \\
        \hline
        Number of  cities & 3 & 4 & 3 &  \\
        Number of trucks & 1 & 1 & 2 &  \\
        Number of qubits for routing & 9 & 16 & 18 &  \\
        Number of qubits for capacities & 2 & 3 & 3 &  \\
        Total number of qubits & 11 & 19 & 21 &  \\
        \hline
    \end{tabular}
\end{table}

\begin{figure*}
    \centering
        \includegraphics[width=\linewidth]{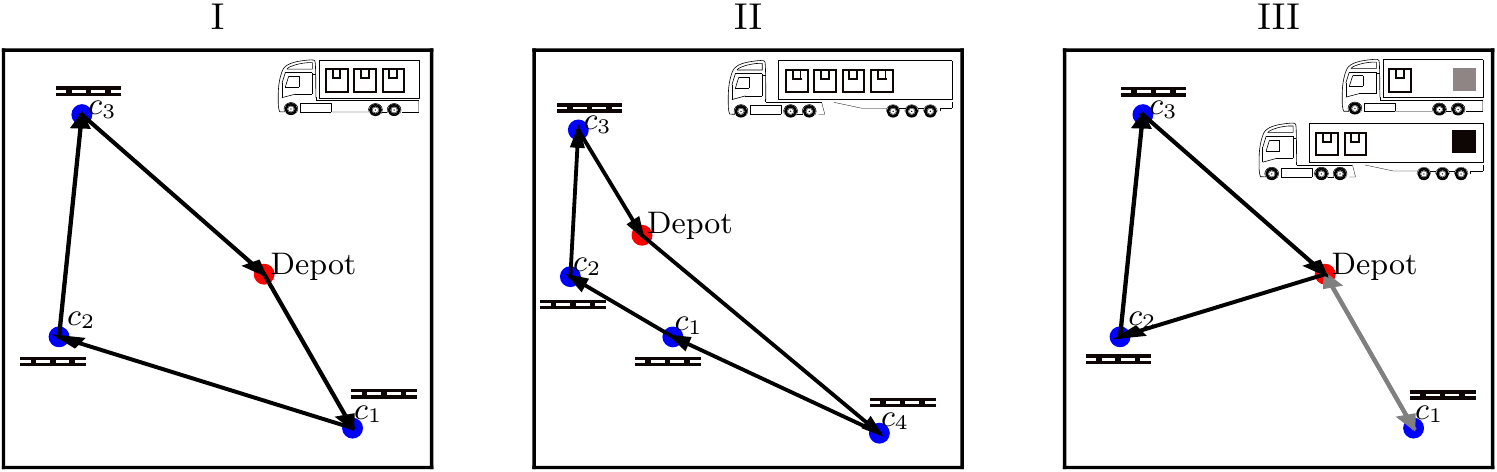}
    \caption{The three problem instances and some solutions for them obtained with low-depth \ac{QAOA}. A visual representation, with correct scaling, of the three problem instances that are considered for the simulations. The optimal solution is shown. Each truck carries a predefined amount of goods and brings it to the respective customers. The pallets indicate that one item has to be carried to the customer. The crates show how much goods is carried by each truck. A color coding indicates the route assignment for problem instance III. The optimal order in which the customers are visited is indicated by arrows (the reverse order is also optimal).}
    \label{fig:problem_instance}
\end{figure*}

In this section, we show numerical results from noise-free simulations of the \ac{QAOA} applied to the \ac{HVRP}. We examine three different problem instances, labelled I, II, and III, which use 11, 19, and 21 qubits, respectively. Table~\ref{tab:problem_instance} contains information about the number of cities, available trucks, and the overall number of qubits needed to encode these problem instances in an Ising Hamiltonian using the scheme we have described in \secref{sec:ising_formulation_for_the_hvrp}. For the simulations, we consider realistic fuel consumption, gas prices, and fixed costs for each truck type, as detailed in Appendix~\ref{appendix}. A graphical representation of the problem instances is shown in \figref{fig:problem_instance}.

For the simulations we consider two different approaches. One is to only solve for satisfying the constraints. The other is to solve the full problem, optimizing not only for feasible solutions, but for the best solution. This stepwise approach, starting with the constraints [Eqs.~(\ref{equ:h_c_routing}), (\ref{equ:h_d_routing}), and (\ref{equ:h_e_knapsack})] and then including the optimization part [(Eqs.~(\ref{equ:h_a_routing})--(\ref{equ:h_b_routing})], helps us understand whether some parts of the full problem contribute more to its difficulty than others.

For the first approach, finding satisfying solutions, we neglect the actual optimization part of the problem, i.e., minimizing the cost of the routing for the solution. We set the prefactors of the different parts of the Hamiltonian [Eqs.~(\ref{equ:h_c_routing}), (\ref{equ:h_d_routing}), and (\ref{equ:h_e_knapsack})] to 1. The eigenvalues of the Hamiltonian are integers. This allows us to restrict the search space for $\beta$ and $\gamma$ in Eqs.~(\ref{equ:h_cost})--(\ref{equ:h_mixing}) to $[0, \pi ]$ and $[ 0, 2 \pi ]$, respectively. For the full problem, we cannot make use of this simplification.

The second approach is to solve the full problem with the optimization included [Eqs.~(\ref{equ:h_a_routing})--(\ref{equ:h_d_routing}) and \eqref{equ:h_e_knapsack}]. Additionally, we rescale the cost function of the optimization [Eqs.~(\ref{equ:h_a_routing})--(\ref{equ:h_b_routing})] such that it only takes values between 0 and 1. Note that for rescaling the cost, we have to evaluate the cost for each possible solution, which makes it necessary to brute-force the problem. This is only feasible for small problem instances. In a real-world application, this rescaling procedure cannot be applied and therefore it might be necessary to optimize over the penalty weights as well~\cite{Roch2020, Svensson2021}. The eigenvalues are not integer values anymore and therefore the entire search space for the variational parameters must be explored.


\subsection{Energy landscape}

\begin{figure*}
    \centering
    \includegraphics[width=\linewidth]{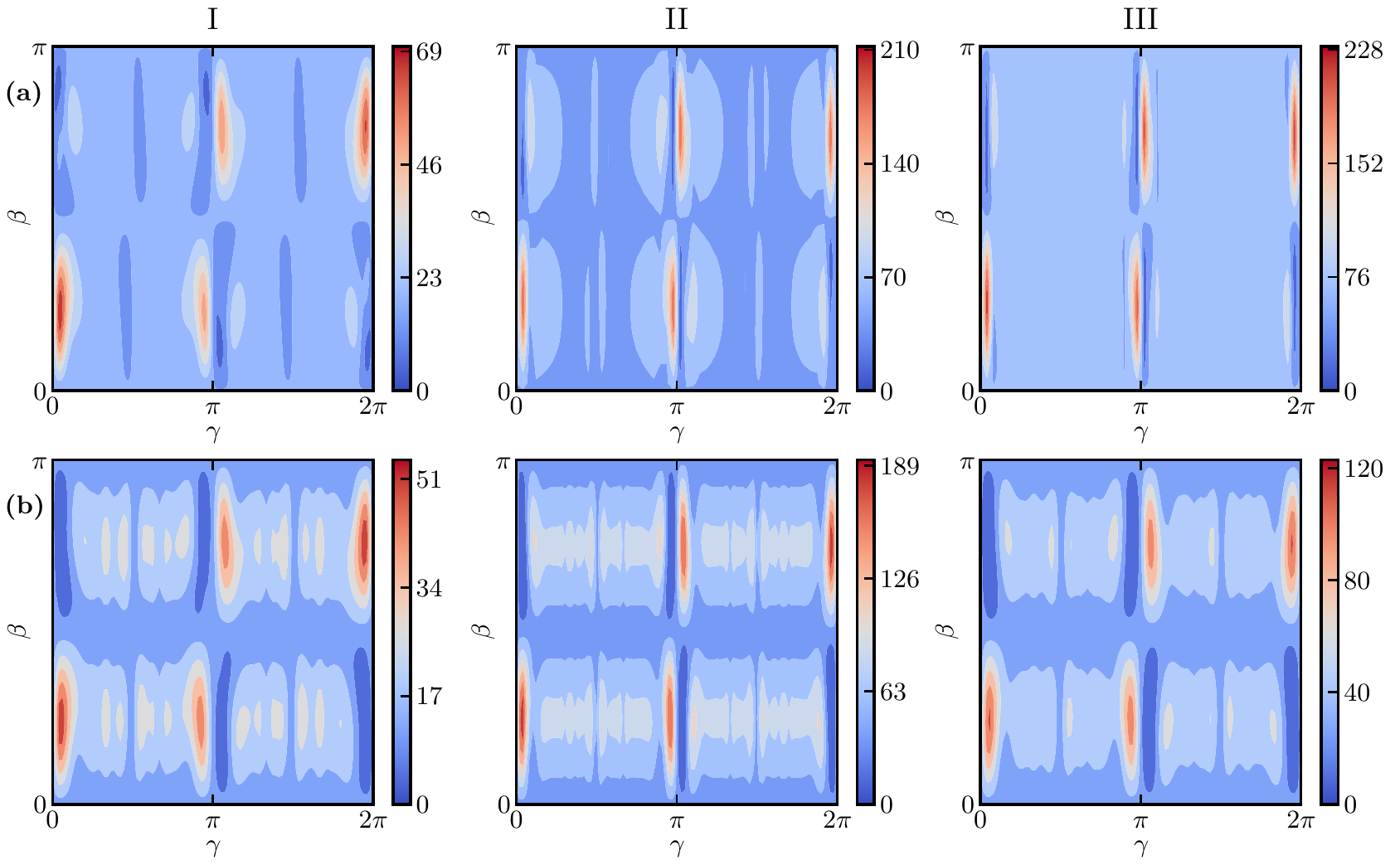}
    \caption{Energy landscapes for the three problem instances with circuit depth $p=1$.
    (a) The energy landscape for the full \ac{HVRP} [Eqs.~(\ref{equ:h_a_routing})--(\ref{equ:h_d_routing}) and \eqref{equ:h_e_knapsack}]. The expectation value $E(\gamma, \beta)$ for the total cost depends on the classically optimized angles $\gamma$ and $\beta$. The periodicity is broken due to the non-integer eigenvalues for the cost Hamiltonian.
    (b) The energy landscape for the capacity constraint only [\eqref{equ:h_e_knapsack}]. Here, the expectation value $E(\gamma, \beta)$ describes the energy penalty for breaking the capacity constraint.
    \label{fig:contour_plots}}
\end{figure*}

To illustrate the difficulty of finding good variational parameters for the \ac{QAOA}, we show in \figref{fig:contour_plots} the energy landscape for $p=1$ for each problem instance considering the full problem [\figpanel{fig:contour_plots}{a}] and the capacity constraint in isolation [\figpanel{fig:contour_plots}{b}]. We evaluate the expectation value $E (\gamma, \beta)$ of the cost Hamiltonian on a grid $\{ \gamma, \beta\} \in [0, 2 \pi ] \times [ 0, \pi ]$. Note that for the full problem, the entire space of variational parameters must be considered, but for this visualization we constrain it to the range mentioned.

The states with the lowest energy are marked with dark blue in \figref{fig:contour_plots}. Each surface plot shows four distinct optima. Moreover, the variational parameters are concentrated in the same region for all three problem instances, both for the full problem and for only the capacity constraint. The range for optimal parameters narrows with increasing problem size. Similar behaviour has been observed in Ref.~\cite{Streif2019}. The overall shape of the energy landscape does not change significantly with varying problem size, but the overall expectation value increases. This is not surprising, since with increasing problem size there are many more constraints to satisfy, goods to deliver, and trucks to choose from. 

As discussed in \secref{sec:ising_formulation_for_the_hvrp}, the \ac{HVRP} consists of two problems, a routing problem and a capacity problem. The capacity problem is analogous to the constraints of the knapsack problem~\cite{Lucas14}. To better understand the energy landscapes for the capacity part of the three problem instances shown in \figpanel{fig:contour_plots}{b}, we now take a closer look at the landscape of a particular knapsack problem. The problem is: given a knapsack with a maximum capacity of 5, choose from the list of items $[4,3,2,1]$ the ones that satisfy the capacity constraint. Seven qubits are needed to encode the problem, making use of the log trick introduced in \secref{reduce_number_of_auxiliary_qubits}.  

\begin{figure}
    \centering
    \includegraphics[width=\linewidth]{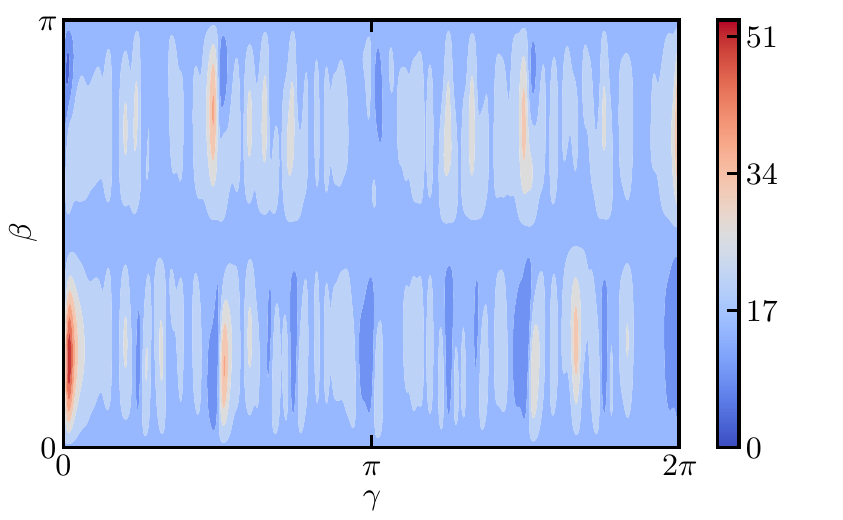}
    \caption{The energy landscape for the knapsack problem [see \eqref{equ:knapsack_constraint}] discussed in the main text with a circuit depth of $p=1$. The energy landscape is highly non-convex and finding the global optimum is difficult for a classical optimizer.}
    \label{fig:energy_landscape_knapsack}
\end{figure}

Figure~\ref{fig:energy_landscape_knapsack} shows the energy landscape for this knapsack problem. The plot shows a rapidly oscillating energy landscape. It is clear that many optimizers will struggle to find the global optimum in this landscape. We argue that with increasing complexity of the problem instances for the \ac{HVRP}, maneuvering the landscape of the capacity constraint becomes a difficult problem. In \figpanel{fig:contour_plots}{b} we do not observe this behaviour yet, but this is simply due to the fact that the problem instances we consider are small (see Table~\ref{tab:problem_instance}). To obtain a landscape that is easier to handle for the classical optimizer, it might be necessary to relax the knapsack constraint or to find a different formulation to encode the capacity constraint~\cite{DelaGrandrive2019}.


\subsection{Increasing the circuit depth}

\begin{figure*}
    \centering
        \includegraphics[width=\linewidth]{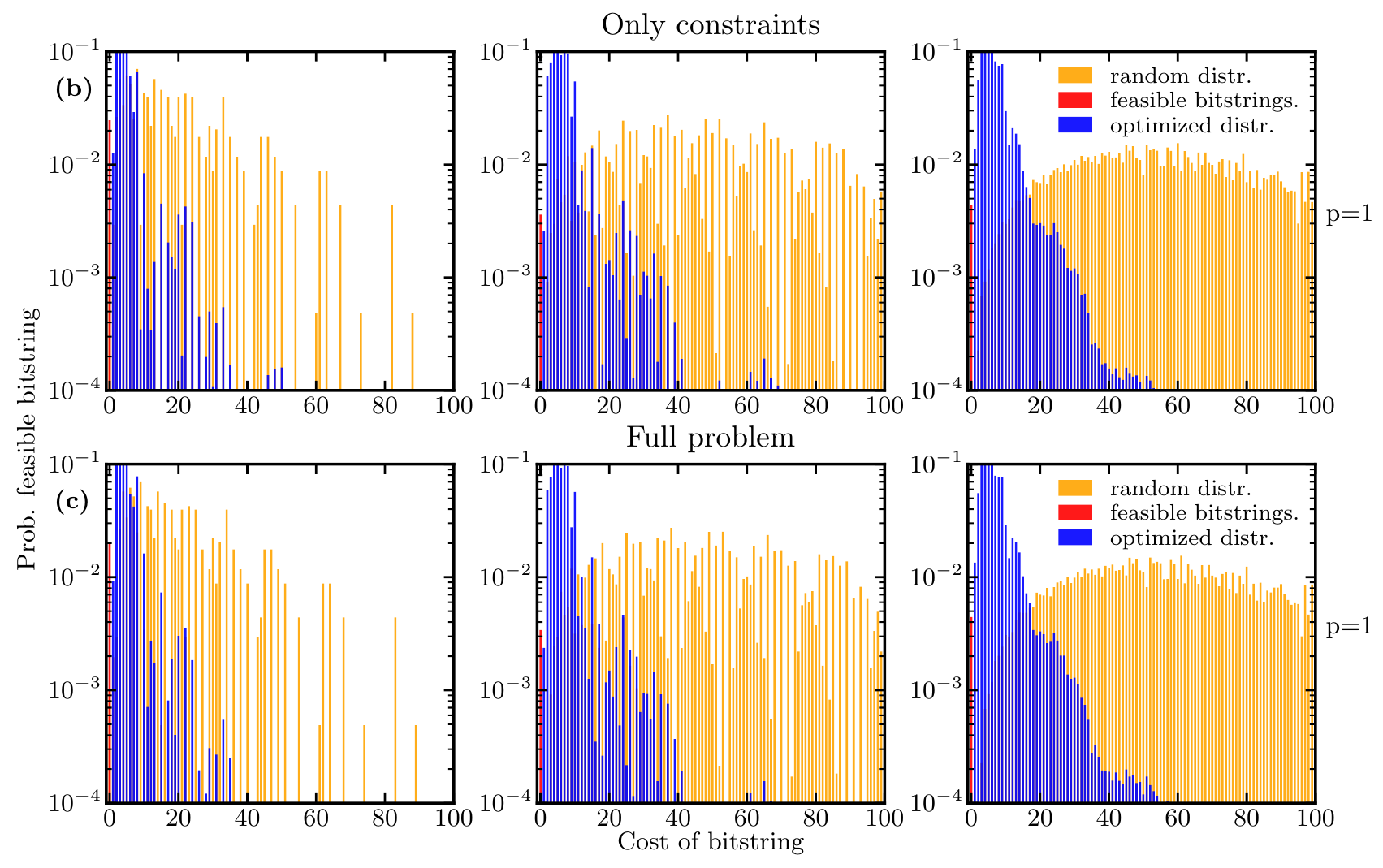}
    \caption{(a) The probability distribution of the optimized variational state (red, blue) for $p=1$. The color red shows the probability for finding a valid solution to the routing problem, i.e., a solution satisfying all constraints. The blue color indicates the overall outline of the optimized probability distribution. As a comparison, we show the probability distribution for a variational state in the $\ket{+}^{\otimes n}$ state, meaning all bitstrings are sampled with uniform probability (orange).
    (b) The probability distribution of the optimized variational state considering the full problem (red, blue) for $p=1$. The probability distribution is binned such that $i$ is the largest possible integer where $i \le x $ holds, also denoted as $\lfloor x \rfloor$. That results in all feasible solutions being binned to zero and all the others being binned to integers indicating the number of constraint violations.
    }
    \label{fig:probability_distribution}
\end{figure*}

It has been shown that for a circuit depth of $p=1$, the \ac{QAOA} cannot outperform classical optimization algorithms~\cite{Farhi2014, Farhi2020}. For actual applications of the \ac{QAOA}, it is therefore necessary to go to a circuit depth beyond $p=1$.

Before we investigate $p>1$, we start with the lowest possible circuit depth of $p=1$. In \figpanel{fig:probability_distribution}{a} and \figpanel{fig:probability_distribution}{b}, we show a histogram of the probability distribution created by the variational circuit for finding solutions satisfying all constraints, in the cases where the cost function encodes only the constraints and the full problem, respectively. We show the probability of sampling bitstrings with a specific cost. Note that there can be several bitstrings leading to a particular cost. The probability of sampling any of the feasible bitstrings is shown in red and the rest of the optimized distribution is depicted in blue. As a comparison, we show the probability distribution for a variational state in the $\ket{+}$ state, meaning all bitstrings are sampled with uniform probability (orange). The difference between these two distributions is marginal. Thus, adding the cost terms [Eqs.~(\ref{equ:h_a_routing})--(\ref{equ:h_b_routing})] to the optimization problem does not alter the overall performance of the algorithm when it comes to satisfying the constraints. This is perhaps not so surprising when considering that due to the rescaling of the cost Hamiltonian, the costs not associated with constraints impact the overall shape of the energy landscape less.

The optimized probability distribution is shifted to the left compared to the random distribution. Thus, sampling from the optimized variational state gives solutions with overall lower energy compared to random sampling of bitstrings. Moreover, for the small problem instance with 11 qubits, it is possible to sample valid bitstrings with a probability of approximately \unit[3]{\%}. For the 19- and 21-qubit problem instances, the probability of sampling a valid bitstring is about an order of magnitude less. This is not surprising since we are limiting the algorithm to a shallow circuit depth.

\begin{figure*}
    \includegraphics{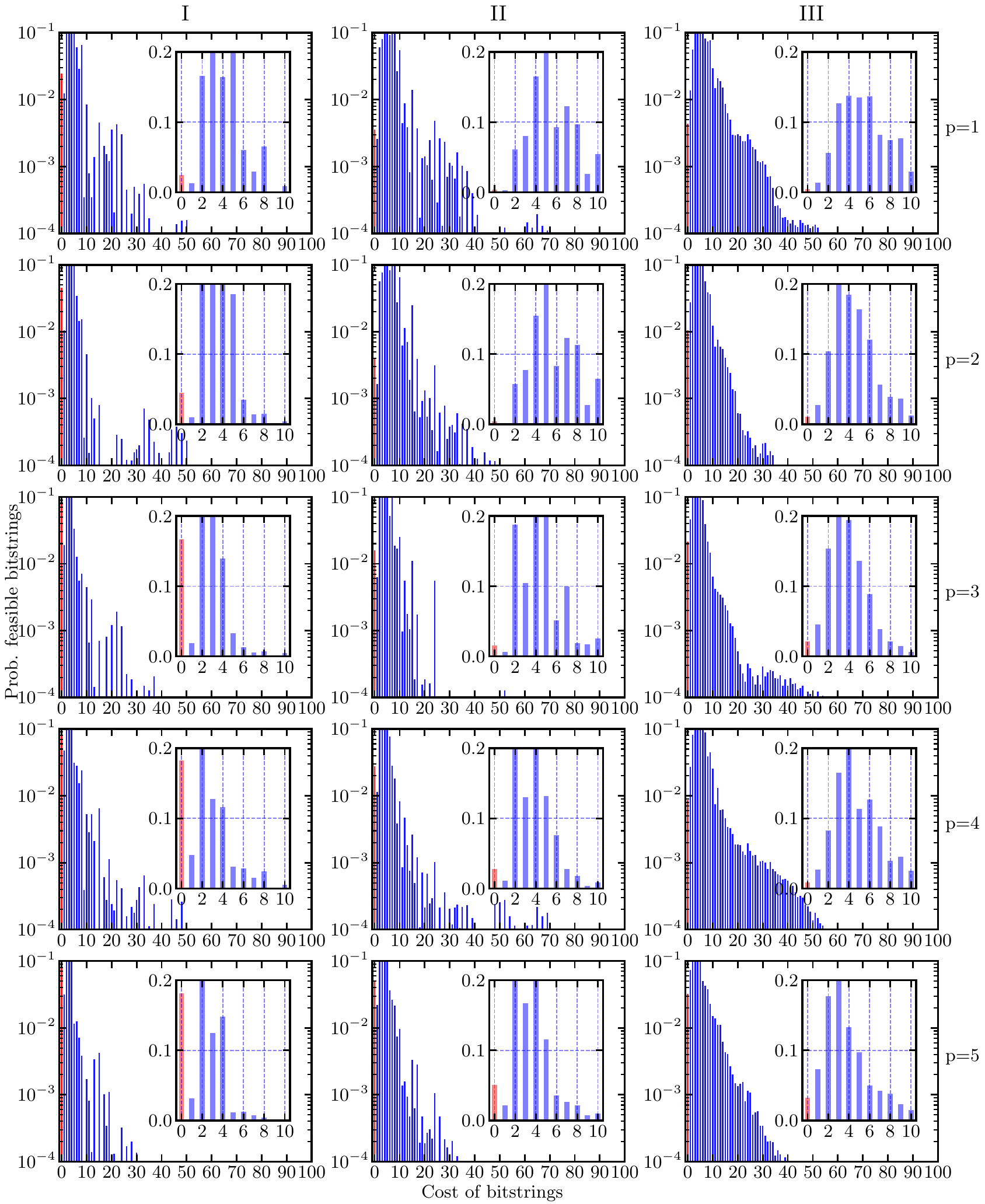}
    \caption{The probability distribution of the variational state $\ket{\boldsymbol{\gamma, \beta}}$ for the 11- (left), 19- (middle) and 21-qubit (right) problem instances as a function of the circuit depth $p$ for obtaining feasible tours. The circuit depth increases from top $p=1$ to bottom $p=5$ and shifts to lower-energy eigenstates with increasing circuit depth. The probability of sampling bitstrings that encode the optimal solution is marked with red. The simulations were conducted with the classical optimizer basinhopping with the local optimizer \ac{BFGS} (see \secref{sec:ComparingOptimizers}).}
    \label{fig:prob_distr}
\end{figure*}

\begin{figure*}
    \includegraphics{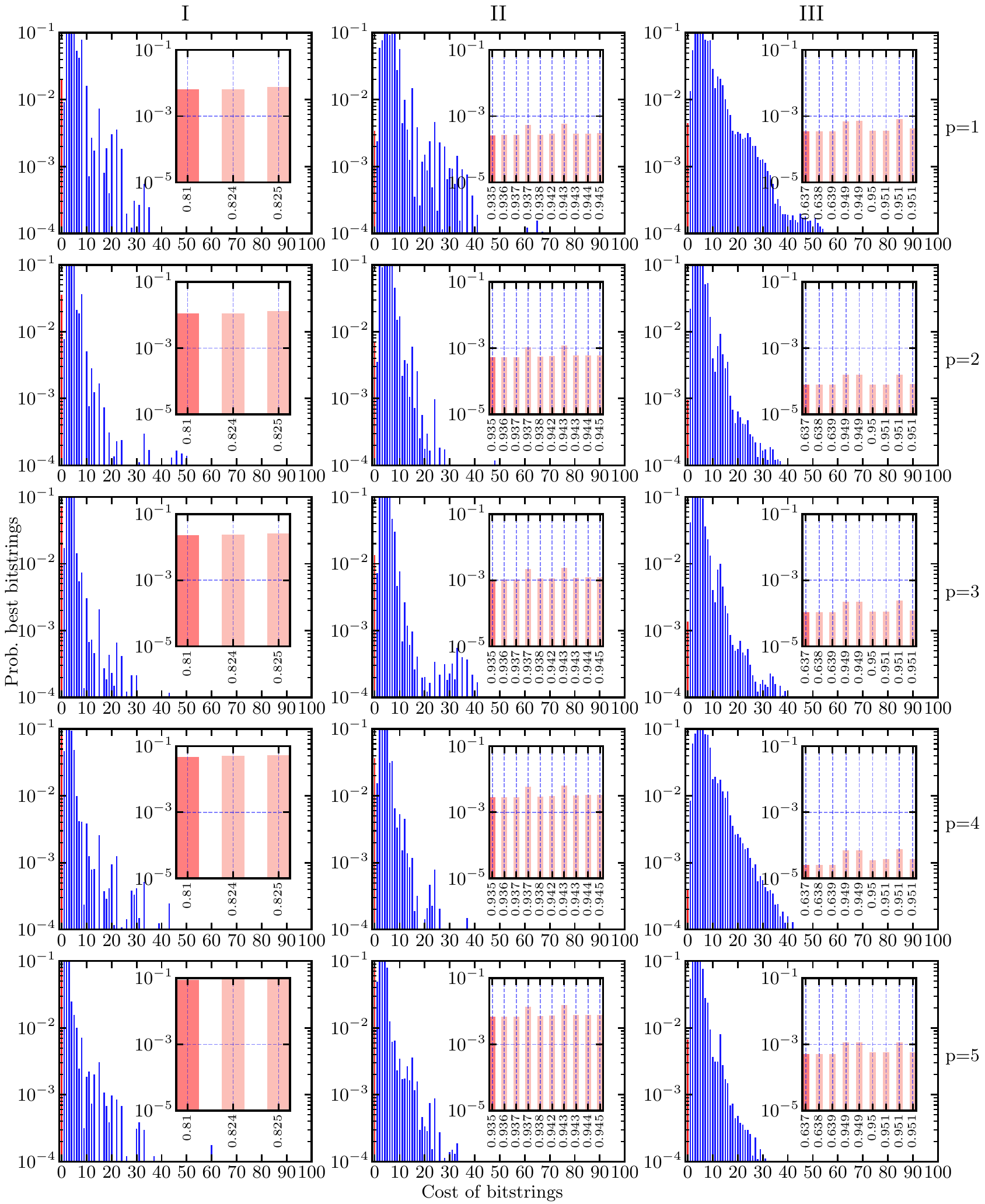}
    \caption{The probability distribution of the variational state $\ket{\boldsymbol{\gamma, \beta}}$ for the 11- (left), 19- (middle), and 21-qubit (right) problem instances as a function of the circuit depth $p$ for finding the best tour. The circuit depth increases from top $p=1$ to bottom $p=5$ and shifts to lower-energy eigenstates with increasing circuit depth. The probability of sampling the best bitstrings is marked with red. The binning is done in the same way as in \figpanel{fig:probability_distribution}{b}. The inset shows the probability for sampling any of the two best bitstrings (dark red, leftmost bin) and the probabilities for sampling any of the other feasible bitstrings (light red).
    The simulations were conducted with the classical optimizer basinhopping with the local optimizer \ac{BFGS} (see \secref{sec:ComparingOptimizers}).}
    \label{fig:prob_distr_full_problem}
\end{figure*}

Next, we go to a circuit depth above $1$. Here we consider in \figref{fig:prob_distr} and \figref{fig:prob_distr_full_problem} finding feasible solutions and the best solutions, respectively. We shows how the variational state changes with increasing circuit depth from $p=1$ to $p=5$ for the three problem instances [see \figref{fig:problem_instance}]. Additionally, we show  for \figref{fig:prob_distr} an inset focusing on the low-energy part of the distribution of the variational state and for \figref{fig:prob_distr_full_problem} an inset focusing on all feasible bitstrings. Again, the bitstrings we are aiming to sample are marked in red and the rest of the optimized distribution is depicted in blue.

For the 11-qubit instance, the probability of sampling a valid bitstring reaches values up to \unit[18]{\%} for $p=5$. For the slightly larger instances (19 and 21 qubits), the probability of sampling a valid bitstring does not exceed \unit[5]{\%} (see \figref{fig:prob_distr}). For the more difficult problem of actually finding the optimal tour (not just a feasible tour), the probability drops to \unit[9]{\%} for the 11-qubit instance, while for the 19- and 21-qubit instances the probability of sampling the ideal bitstring is below \unit[1]{\%} (see \figref{fig:prob_distr_full_problem}).

The inset in \figref{fig:prob_distr_full_problem} shows that the algorithm cannot distinguish between the different feasible solutions and thus fails to optimize for the best solution. This might be due to the very small energy gap between the lowest and next lowest energy eigenstate, which is an artifact of the rescaling of the cost we introduced earlier. The point of this rescaling is to ensure that all feasible solutions have lower energy than any solution violating any constraint. Rigorous hyperparameter optimization might be necessary to weight the cost and the constraints to circumvent the problems introduced by the chosen rescaling~\cite{Roch2020}. 

The probability of sampling bitstrings with low cost increases with increasing circuit depth, meaning that the probability distribution shifts to lower-energy eigenstates. This trend is visible in~\figref{fig:prob_distr} and \figref{fig:prob_distr_full_problem} as the probability distribution shifts increasingly to the left with increasing depth. It might be possible with increasing circuit depth to obtain higher probabilities of sampling optimal solutions at the expense of optimizing more variational parameters. This is in accordance with the theory of \ac{AQO} --- the performance of the algorithm becomes better with more variational parameters. The drawback is that the optimization problem becomes increasingly difficult and time-consuming. It is therefore important to select a good optimization method.


\subsection{Performance comparison of classical optimizers}
\label{sec:ComparingOptimizers}

Optimizers for finding the variational parameters play an important role in the context of \ac{VQAs}. Much of the current research aims to find optimizers that perform well on these quantum circuits~\cite{Barkoutsos2020, Kubler2020, Spall1992, verdon2019learning}. There is a rich literature on optimizers for variational circuits proposing different optimizers for different problems~\cite{Cerezo, Lavrijsen2020}.
In the following, we analyze the performance of simulations using four different well-known optimizers: Nelder-Mead~\cite{Nelder1965}, Powell~\cite{powell1964}, differential evolution~\cite{Storn1996}, and basinhopping~\cite{Wales1997}. The selected optimizers work differently: some use global search mechanisms consisting of multiple random initial guesses while others use only a single random initial guess as a starting point for the optimization. These characteristics are crucial to understand why the performance of the optimizer varies.

The Nelder-Mead method uses a geometrical shape called a simplex to search the function space. 
With each step of the optimization, the simplex shifts, ideally, towards the region with a minimum. The Nelder-Mead algorithm belongs to the class of gradient-free optimizers.
The Powell optimzer works for non-differentiable functions; no derivatives are needed for the optimization. The method minimises the function by a bi-directional search along each search vector. The initial search vectors are typically the normals aligned to each axis. 
The differential evolution algorithm is stochastic in nature and does not rely on derivatives to find the minimum. This algorithm often requires larger numbers of function evaluations than conventional gradient-based techniques. 
The basinhopping optimization algorithm is a two-phase method, which couples a global search algorithm with a local minimization at each step. For the simulations here (including in the preceding sections), we used the \ac{BFGS} algorithm~\cite{FredericBonnans2006} as a local optimizer. This framework has been proven useful for hard nonlinear optimization problems with multiple variables~\cite{Olson2012}.

To speed up the optimization routines, we use the optimized parameters from $p-1$ as an intial guess for the optimization of the variational circuits for $p>1$, as well as the optimized parameters from the 11-qubit problem instance as an initial guess for the 19- and 21-qubit instances. The concentration of variational parameters for the \ac{QAOA} has been observed by many researchers~\cite{Streif2019, Akshay2021}. It has been shown for a circuit depth of $p=1,2$ that the variational parameters for small problem instances can be used to infer parameters for larger problem instances~\cite{Akshay2021}. This can significantly speed up the optimization of the \ac{QAOA}. 

\begin{figure*}
    \centering
    \includegraphics[width=\linewidth]{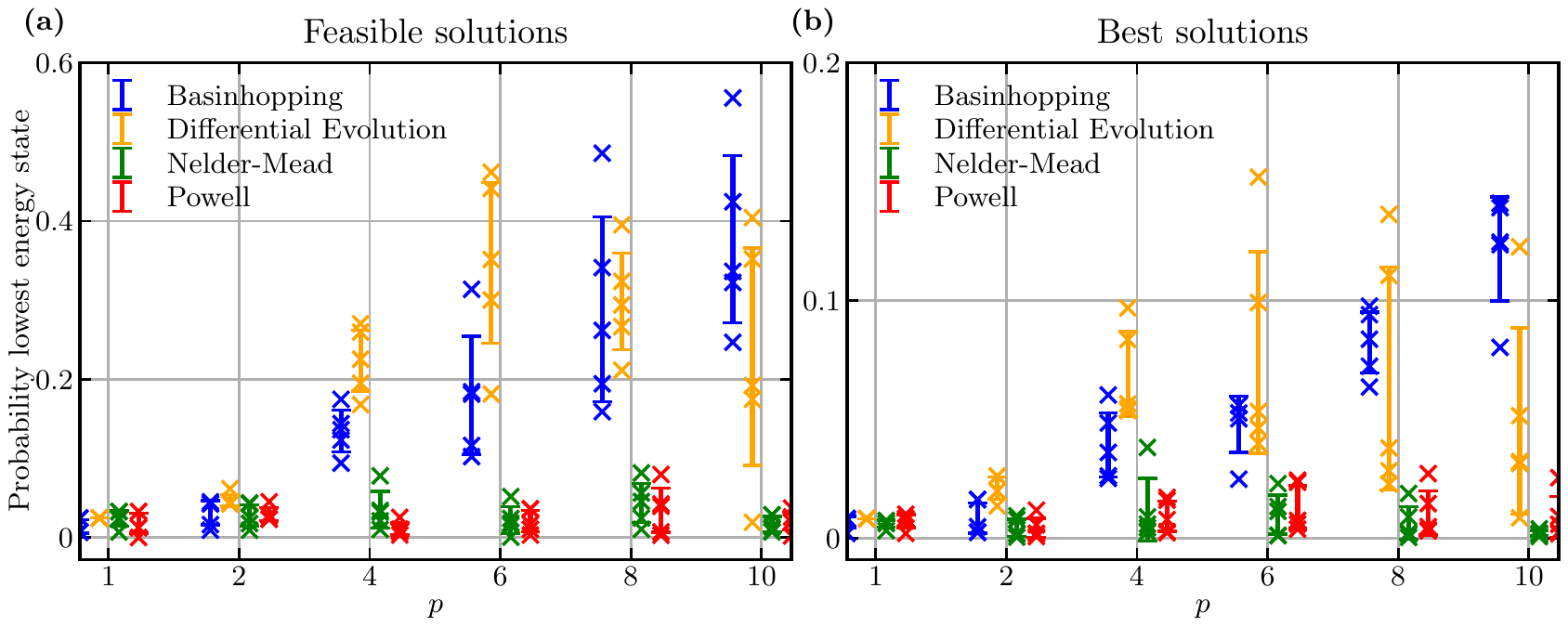}
    \caption{Comparing different optimizers for the \ac{HVRP} on the 11-qubit problem instance shown in \figref{fig:problem_instance}. A probability of 1 means that a feasible bitstring is always sampled. 
    (a) Success probability for finding a valid solution with increasing circuit depth $p$.
    (b) Success probability for finding the best solution with increasing circuit depth $p$.
    \label{fig:compare_optimizer}}
\end{figure*}

In \figpanel{fig:compare_optimizer}{a}, we show the success probability of finding a viable bitstring as a function of the circuit depth $p$ for the 11-qubit problem instance. Similarly, we show the success probability of sampling one of the two optimal tours as a function of the circuit depth $p$ in \figpanel{fig:compare_optimizer}{b}. One of the optimal tours is shown in \figref{fig:problem_instance}. The other optimal solution is following the same path in reverse order.

The results show in \figref{fig:compare_optimizer} clearly favours the optimizers basinhopping and differential evolution. They perform significantly better than the Nelder-Mead and Powell optimization algorithms.
This difference in performance might be due to the global optimization routine that these latter two algorithms use to find the optima. The performance of the Nelder-Mead algorithm strongly depends on the initial simplex and the simplex is usually randomly generated. Depending on the starting point, the performance can vary, but it usually cannot compete with the solution quality of the basinhopping or differential-evolution algorithms. 


\subsection{Runtime comparison of the classical optimizers}

\begin{figure}
    \centering
    \includegraphics[width=\linewidth]{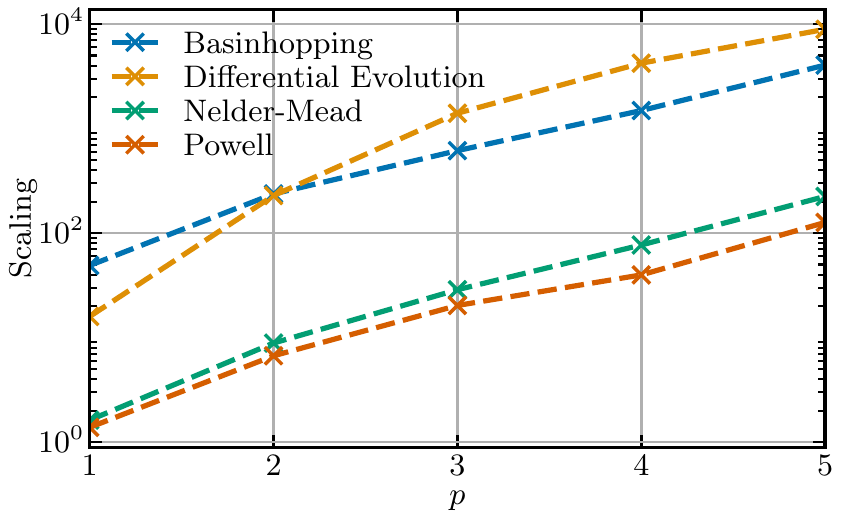}
    \caption{Runtime comparison for the four classical optimizers tested in this work. The plot shows the runtime as a function of circuit depth $p$. 
    \label{fig:time_comparison}}
\end{figure}

For applications it is also important to understand the scaling of the runtime of the optimization routine with increasing circuit depth. Here we benchmark the classical optimizers from the preceding subsection on this measure. The result is shown in \figref{fig:time_comparison}.

We see that the optimizers basinhopping and differential evolution need roughly one order of magnitude more time for the optimization than the Nelder-Mead and Powell algorithms. This is due to the many circuit queries the former optimizers have to perform. The tradeoff between the runtime of the algorithm and its performance becomes evident, as the slowest optimizers in terms of runtime performed best in terms of success probability (see \figref{fig:compare_optimizer}). Moreover, the linear increase for all optimizers on the semi-log scale in \figref{fig:time_comparison} indicates that the amount of time needed for the optimzation scales exponentially with $p$. However, further investigation is needed to confirm this statement. It might become a difficult problem to tackle when a circuit depth beyond $p=20$ is considered.

As mentioned earlier, researchers are already investigating how the optimization could be simplified or circumvented completely~\cite{Streif2019, Akshay2021}. The results here further underlines the importance of such research.


\section{Conclusion}
\label{sec:conclusion}

We have derived an Ising formulation for the heterogeneous vehicle routing problem (\ac{HVRP}) under consideration of all relevant constraints, enabling this problem to be solved on a quantum computer. In our formulation, the number of qubits needed to encode a problem instance scales quadratically with the number of customers. Therefore, quantum computers will need to have at least millions of qubits to use our suggested encoding scheme to solve problem instances that are at the limit of what today's classical high-performance optimizers can solve. A quantum advantage could still be had with fewer qubits for smaller problems if they could be solved faster on a quantum computer than on a classical one, but the present work did not give indications of such speed-ups for small problems.

We simulated solving small instances of the \ac{HVRP} and with the quantum approximate optimization algorithm (\ac{QAOA}). We considered three distinct problems, requiring 11, 19, and 21 qubits, respectively. We investigated the performance of the algorithm with respect to two design choices: the classical optimizer and the depth $p$ of the quantum circuit.

For the choice of optimizer, we found that the basinhopping and differential-evolution algorithms seem well suited to optimize the variational parameters of the quantum circuit. However, this performance came at the expense of comparably long optimization times. Furthermore, our data indicates that the optimization time needed to find suitable angles for the variational quantum circuit increases exponentially with $p$, but further investigation is needed to verify this scaling.

We have seen that routing with additional capacity constraints is a difficult problem for the hybrid quantum-classical approach to handle. The problem becomes more evident when isolating the inequality constraint. Then we can see that the energy landscape has multiple scattered local minima. In future work, one might want to consider a different formulation for the capacity constraint~\cite{DelaGrandrive2019}.

Moreover, \figref{fig:prob_distr_full_problem} shows that the \ac{QAOA} fails to distinguish solutions that satisfy all constraints but differ in cost. This failure may be due to the small energy gap between the different feasible solutions and could be circumvented by restating the problem such that feasible solutions are separated by larger energy gaps. One way of achieving this goal is to conduct a rigorous hyperparameter optimization to weight the cost and the constraints accordingly. Similar ideas were explored for the knapsack problem in Ref.~\cite{Roch2020}.

\section{Outlook}

The classical optimization procedure is a central component in all variational algorithms and key to their success. Therefore, finding new optimizers is one interesting area of research. Several works have investigated machine-learning-based optimizers~\cite{verdon2019learning, yao2020policy, garciasaez2019quantum}. A study of such optimizers for the \ac{HVRP} could boost the performance of the algorithm.

Even though we assumed a noise-free system, the performance is not competitive with modern high-performance heuristics~\cite{ortools, Uchoa2017a} which can solve instances with more than 1,000 customers. Such optimizers are not limited in the number of decision variables, but rather by the running time for the optimization. However, the problems we considered are too small for a meaningful comparison. Furthermore, the assumption of a noise-free system does not hold for \ac{NISQ} computers and a decrease in performance can be expected if the algorithm is executed on such a device~\cite{Preskill2018, Streif2020}. An interesting topic for future work would be to investigate the \ac{HVRP} in combination with \ac{QAOA} under the assumption of a noisy system. Moreover, running small problems on an actual quantum computer could give a better view on the applicability of the \ac{QAOA} to the \ac{HVRP} and its competitiveness with classical heuristics.

In \figref{fig:energy_landscape_knapsack}, we could see that the knapsack constraint creates an energy landscape with rapidly oscillating local minima. This makes it difficult for many optimizers to find a good approximate solution. It would be interesting to investigate why this particular problem is difficult and if it is possible to relax the knapsack constraint or reformulate it such that the optimization landscape becomes easier to maneuver.

Furthermore, the \ac{QAOA} can be expanded to the quantum alternating operator ansatz~\cite{Hadfield2019}. Investigating different mixer Hamiltonians for the \ac{HVRP} could lead to a better overall performance. Ideally, the mixer Hamiltonian would provide a framework that keeps the algorithm in the subspace of allowed solutions~\cite{Wang2019}. Currently, the standard mixer Hamiltonian used in this work makes the algorithms explore every possible bitstring.

Finally, we note that for real-world applications it might be necessary to consider multiple depots. Therefore, another avenue to explore are more Ising formulations that allow for solving different variations of VRPs. 

\begin{acknowledgments}

DF, MG, and AFK acknowledge financial support from the Knut and Alice Wallenberg foundation through the Wallenberg Centre for Quantum Technology (WACQT). Computations were performed on the Vera cluster at Chalmers Centre for Computational Science and Engineering (C3SE).

\end{acknowledgments}

\appendix
\section{Additional information for the simulation of the \ac{QAOA}}
\label{appendix}

The information used to determine the fixed and variable cost for the \ac{QAOA} simulations in this work is shown in  \tabref{tab:fixed_variable_cost}. The table shows the fuel consumption and cost per \unit[100]{km}, the price of the chassis, and loading capacity~\cite{toheed} for the two truck types we consider in \figref{fig:problem_instance}: a rigid truck (\texttt{rt}) and a tractor–semitrailer (\texttt{ts}).

\begin{table}
	\centering
	\caption{Fuel consumption $f$ and corresponding cost $c_{\rm road}$ per 100km, the price $c_{\rm chass}$ of the chassis as well as the loading capacity~\cite{toheed} for each of the two different truck types considered in the simulations of this paper. Note that the loading capacity does not take into account the size or weight of items that can be carried by the truck. The information in this table is used to determine the variable and fixed costs in Eqs.~\ref{equ:h_a_routing} and \ref{equ:h_b_routing}.}
	\label{tab:cost_variable}
	\begin{tabular}{l c c}
		\hline \hline
		    & \texttt{rt} & \texttt{ts}  \\
		\hline
	    $f$ [l/\unit[100]{km}] & 28.6 \cite{site} & 34.5 \cite{semitrailer}  \\
	    $c_{\rm road}$ [\euro/\unit[100]{km}]\footnote{The price per liter is taken to be \unit[1.2]{\euro}~\cite{fuel_price}.} & 34.32 & 41.4\\
	    $c_{\rm chass} [$\euro$]$ & 75,000  & 150,000 \\
		$m_{\rm max}$ [items] & 3 & 4\\
		\hline \hline
	\end{tabular}
	\label{tab:fixed_variable_cost}
\end{table}

\bibliography{Bib}

\end{document}